\documentclass[aip,pop,preprint]{revtex4-1}
\usepackage{graphicx}
\usepackage{xcolor}
\usepackage{dcolumn}
\usepackage{bm}
\usepackage{float}

\begin{document}

\title{Elliptical Magnetic Mirror generated via Resistivity Gradients for Fast Ignition ICF}
\author{A.P.L.Robinson}
\email{alex.robinson@stfc.ac.uk}
\affiliation{Central Laser Facility, STFC Rutherford-Appleton Laboratory,Didcot, OX11 0QX, United Kingdom}
\author{H.Schmitz}
\affiliation{Central Laser Facility, STFC Rutherford-Appleton Laboratory,Didcot, OX11 0QX, United Kingdom}
\begin{abstract}
The elliptical magnetic mirror scheme for guiding fast electrons for Fast Ignition proposed by Schmitz (H.Schmitz et al., {\it Plasma Phys.Control.Fusion},{\bf 54} 085016 (2012)) is studied for conditions on the multi-kJ scale which are much closer to full-scale Fast Ignition.  When scaled up, the elliptical mirror scheme is still highly beneficial to Fast Ignition.  An increase in the coupling effiency by a factor of 3--4 is found over a wide range of fast electron divergence half-angles.
\end{abstract}
\maketitle

\section{Introduction}
In this paper we make a first assessment of a fast electron guiding concept proposed by Schmitz et al. \cite{schmitz1} which exploits resistivity gradients \cite{robinson1} to self-generate an elliptical magnetic mirror which in turn helps collimate a beam of fast electrons with a large angular divergence.  The original proposal by Schmitz only considered conditions, spatial, and temporal scales that were considerably different to full scale Fast Ignition inertial fusion.  Here we attempt to bridge this disparity and thus make an improved assessment of the concepts utility in Fast Ignition.

Fast Ignition (FI) \cite{tabak1,tabak2} is a variant of inertial confinement fusion (ICF) in which the compression and ignition of the target take place in two separate stages.  The ignition of the target is done by using an extremely high power driver to produce a beam of energetic particles (this could be multi-MeV electrons or ions) to generate a hot spot in the compressed fuel.  Compared to central hot spot ignition, it has the advantages of higher gain for lower total laser energy (a total energy of a few hundred kJ may be possible \cite{atzeni1}).

FI concepts based on fast electrons usually make use of a reentrant cone to shield the path of the ultra-intense laser driver from coronal plasma.  There will be a considerable stand-off distance between the apex of the cone and the centre of the compressed fuel.  A crucial element of the scheme is then the efficient, i.e. well collimated, transport of the fast electrons from the cone apex to the hot spot.  The radius of the laser spot and the hot spot are likely to be comparable, but both are smaller than the stand-off distance by a factor of a few, so the beam must be fairly well collimated to ensure efficient coupling.  If the angular spread in the fast electrons is small, then the resistive generation  of a collimating magnetic field may be sufficient to pinch the beam and provide such collimation.  Considerable effort has been put into modelling the fast electron transport aspect of FI, some of which can be found in \cite{honrubia1,johzaki1,johzaki2,solodov1,solodov2,atzeni3,nicolai1} and references therein.

Extensive experimental and simulation studies seem to indicate that the angular spread of the fast electrons under conditions relevant to both FI and laboratory-scale condtions may be quite large \cite{honrubia1,lancaster1,green1,debayle1,stephens1,norreys1}.  This means that any resistive magnetic field generation in the DT fuel will be insufficient \cite{bell1} to significantly pinch the beam, and low coupling into the hot spot will result.  The problem of `fast electron divergence' has motivated a considerable research effort into modifications of the reentrant cone FI concept, and a number of approaches to addressing this problem have been put forward \cite{robinson1,campbell1,nakamura1,kodama1,robinson3,scott1}.

Here we show that the Elliptical Magnetic Mirror concept proposed by Schmitz can be highly beneficial in terms of mitigating the loss of coupling efficiency caused by large fast electron divergence even when one extends one's considerations to conditions much closer to full-scale FI than the original description of the concept.

\section{Elliptical Mirror Concept}
\label{ellipmir}
The elliptical mirror concept\cite{schmitz1} is based on the idea of magnetic collimation, first proposed by Robinson and Sherlock\cite{robinson1}, and later developed in further theoretical and experimental \cite{kar1,ramak1} work.  The electrons are collimated by azimuthal magnetic fields which are generated by resistivity gradients inside the target. One can assume both charge and current neutrality, i.e. $j_f +j_b = 0$, where $j_f$ is the fast electron current and $j_b$ is the background electron current. Calculating the background current through Ohm's law, $\eta j_b = E$, where $\eta$ is the conductivity, Faraday's law can be written as\cite{jrd1}
\begin{displaymath}
\frac{1}{c}\frac{\partial \mathbf{B}}{\partial t} = \eta\nabla\times\mathbf{j}_f + 
\left(\nabla\eta\right)\times\mathbf{j}_f
\end{displaymath}
The first term on the right-hand side generates a magnetic field that directs electrons towards regions of higher current density, and thus acts to collimate the fast electron beam. The second term, which forms the basis of the elliptical mirror concept, generates a magnetic field at resistivity gradients which acts to keep the fast electrons within regions of higher resistivity. The resistivity gradient is created by a transition between two materials with different $Z$. The high energy electrons are generated within the high-$Z$ material and the magnetic field at the material interface will deflect the electrons and keep them inside the high-$Z$ domain. 

For sufficiently steep resistivity gradients one can assume that the electrons are reflected specularly off the magnetic layer. In this case the electrons feel the magnetic field only for a short time and move only a short distance along the tangential direction. If the magnetic field in the layer does not vary along the tangential direction over this short distance then the electron trajectory is symmetric with respect to the normal direction and the electrons are specularly reflected. This can be used to construct geometries for collimating electrons similar to optical mirrors. In this approximation, an elliptical geometry with one focal point in the injection region and the other focal point in the hot spot should be highly effective at focusing the electron beam into a hot spot of roughly the same size as the laser spot.

\section{Simulations}
\subsection{Set Up}
\begin{figure}[H]
\begin{center}
\includegraphics[width = \columnwidth]{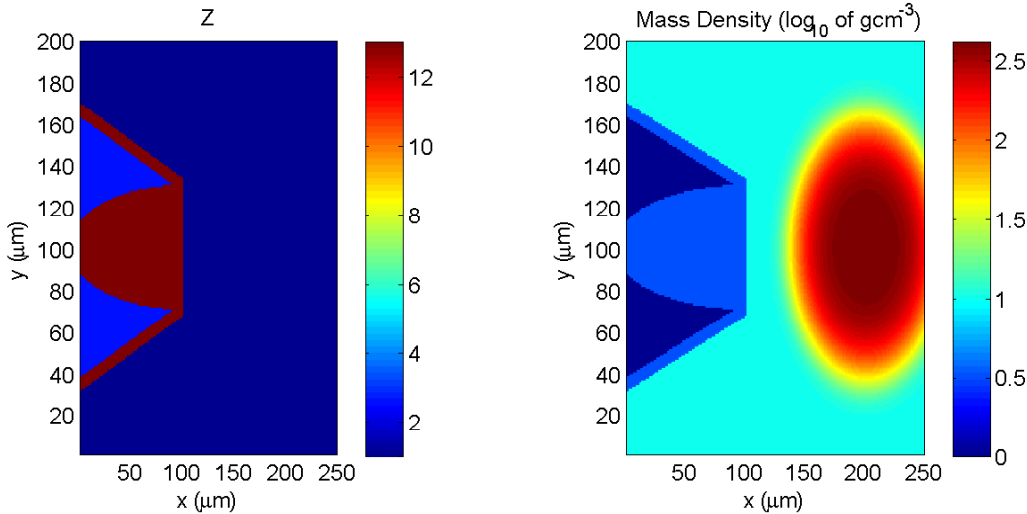} 
\caption{\label{fig:figure1} Target set-up in baseline simulation.  Left : Plot of target $Z$.  Right : $log_{10}$ plot of target mass density (in gcm$^{-3}$). }
\end{center}
\end{figure}

Simulations were performed using the 3D particle hybrid code {\sc zephyros} \cite{robinson2,kar1,ramak1}.  This is based on the hybrid method developed by Davies in a series of publications \cite{jrd1}.  A 250$\times$200$\times$200 grid was used with a 1$\mu$m cell size in each direction.  The target set up is shown (in terms of atomic number and mass density) in figure \ref{fig:figure1}, and these profiles are axisymmetric about the $y=z=$100$\mu$m line.  The target consists of Al re-entrant cone the top 100 $\mu$m of which contains a guiding structure consisting of a truncated semi-ellipsoid core of Al surrounded by a CH$_2$ substrate.  Outside the core is a mass of compressed DT, the centre of which is centred at ${\bf r}_{DT} =$(200,100,100)~$\mu$m.  The radial density of the fuel profile is $\rho = 10 + 400\exp\left[-(r/R)^4\right]$gcm$^{-3}$, with $R =$50$\mu$m.  The left hand edge of the simulation domain represents the edge of the insert in the cone tip.  To the left of this there will only be vacuum, and the laser irradiates this surface.

The background temperature is initially set to 100eV everywhere.  The background resistivity was described by the model of Lee and More.  The temporal profile of the injected fast electron beam is a top-hat function of 18~ps duration, and the transverse profile is $\propto \exp\left[-(r/r_{spot})^4\right]$ with $r_{spot} =$15$\mu$m.  The injected fast electron beam models irradiation at an intensity of 4$\times$10$^{20}$Wcm$^{-2}$, with the assumption of 50\% conversion efficiency.  The total injected fast electron energy is 23.00~kJ.  The fast electron distribution is of the form $\propto \exp(-\varepsilon/T_f)$ with $T_f = T_{pond}$ (where $T_{pond}$ is the temperature obtained from Wilk's ponderomotive scaling).  The actual value of $T_f$ being 2.6~MeV.  The angular distribution of the fast electrons is a uniform distribution over a cone subtended by an half-angle of $\theta_{div}$, where  $\theta_{div}=$70$^\circ$ in the baseline case.  The main parameters varied in these simulations (designated A--G) were $\theta_{div}$, and $r_{spot}$.  We also considered the case of a combination of higher intensity and sub-ponderomotive fast electron temperature scaling.

A number of control runs (designated CMP1--CMP4) were also carried out to examine the case where the elliptical mirror is not present.  In the control simulation only the cone tip and the compressed fuel were present in a 200$\times$200$\times$200 box.  All other simulation parameters were same as the base-line.  A table summarizing the parameters used in the different runs is shown below (table \ref{param_tbl}).

The geometry of the Al semi-ellipsoid is chosen so that the centre of the injection region sits at one of the foci of the (full) ellipsoid, and so that the target `hot-spot' sits at the other focus.  As can be seen from figure \ref{fig:figure1}, the minor axis of the ellipsoid, $b$, is 30~$\mu$m.  The major axis, $a$, was therefore chosen to be 104.4~$\mu$m (from $f = \sqrt{a^2-b^2}$), so that the foci of the ellipsoid sit at the desired positions.  The quality of guiding obtained by this configuration rests on both establishing a strongly reflecting magnetic field at the ellipsoid surface, and on the aberration that arises from having an extended source being negligible.

\begin{table}
\resizebox{\columnwidth}{!}{
\begin{tabular}{|c|c|c|c|c|}
\hline
Simulation & $\theta_{div}$ & $r_{spot}$ ($\mu$m) & $T_f$ & $I_L$ (10$^20$Wcm$^{-2}$) \\
\hline
A & 70$^\circ$ & 15$\mu$m & $T_{pond}$ & 4  \\
\hline 
B & 50$^\circ$ & 15$\mu$m &  $T_{pond}$ & 4  \\
\hline
C & 60$^\circ$ & 15$\mu$m & $T_{pond}$  & 4  \\
\hline
D & 80$^\circ$ & 15$\mu$m &  $T_{pond}$ & 4  \\
\hline
E & 90$^\circ$ & 15$\mu$m & $T_{pond}$ & 4   \\
\hline
F & 50$^\circ$ & 14$\mu$m & $T_{pond}$ & 4 \\
\hline
G & 50$^\circ$ & 13$\mu$m & $T_{pond}$ & 4 \\
\hline
H & 70$^\circ$ & 15$\mu$m & $0.6T_{pond}$ & 6 \\
\hline
I & 60$^\circ$ & 15$\mu$m & $0.6T_{pond}$ & 6 \\
\hline
J & 50$^\circ$ & 15$\mu$m & $0.6T_{pond}$ & 6 \\
\hline
K & 40$^\circ$ & 15$\mu$m & $0.6T_{pond}$ & 6 \\
\hline
CMP1 & 70$^\circ$ & 15$\mu$m & $T_{pond}$ & 4 \\
\hline
CMP2 & 60$^\circ$ & 15$\mu$m & $T_{pond}$ & 4 \\
\hline
CMP3 & 80$^\circ$ & 15$\mu$m & $T_{pond}$ & 4 \\ 
\hline
CMP4 & 50$^\circ$ & 15$\mu$m & $T_{pond}$ & 4 \\
\hline
CMP5 & 50$^\circ$ & 15$\mu$m & $0.6T_{pond}$ & 6 \\
\hline
\end{tabular}
}
\caption{\label{param_tbl}Table of simulation parameters.}
\end{table}

\subsection{Baseline Simulation}
\begin{figure}[H]
\begin{center}
\includegraphics[width = \columnwidth]{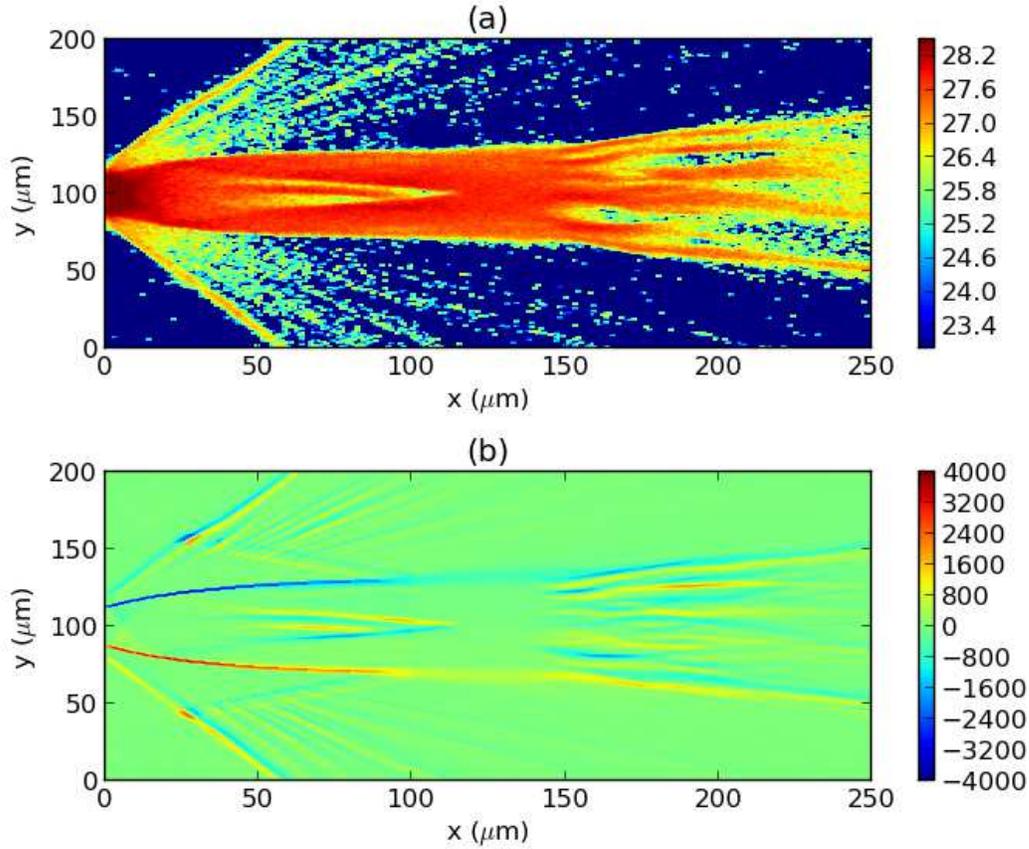} 
\caption{\label{fig:figure2} (a)log$_{10}$ fast electron density (in m$^{-3}$) at 4~ps in baseline simulation (A) in $x$-$y$ mid-plane of simulation box. (b) $B_z$(T) component of magnetic field at 4~ps in baseline simulation (A) in $x$--$y$ mid-plane of simulation box.}
\end{center}
\end{figure}

In the baseline simulation we can see a number of the general features by looking at the magnetic field ($B_z$ component) and fast electron density in the $x$--$y$ mid-plane some way into the simulation.  These are shown at 4~ps in figure \ref{fig:figure2}.

\begin{figure}[H]
\begin{center}
\includegraphics[width = \columnwidth]{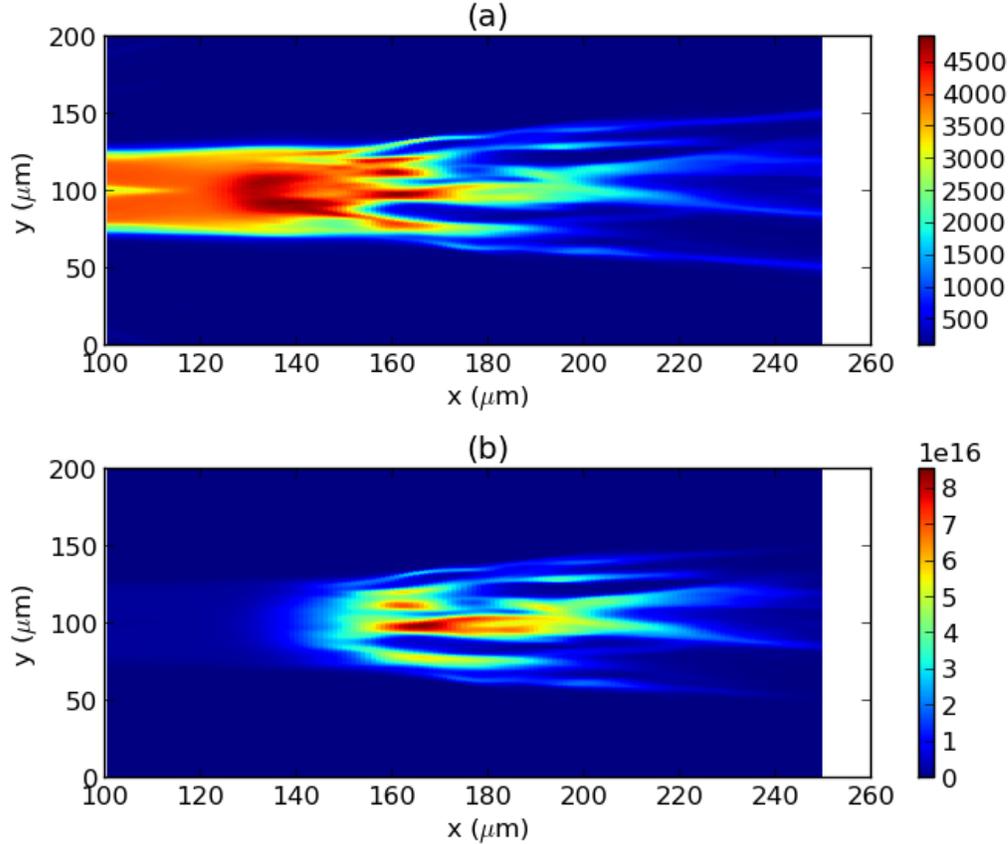} 
\caption{\label{fig:figure3}(a) Ion temperature at 20~ps in eV in baseline simulation (A) for $x >$ 100$\mu$m, (b)
Ion internal energy density at 20~ps in Jm$^{-3}$ in baseline simulation (A) for $x >$ 100$\mu$m.}
\end{center}
\end{figure}

Fig. \ref{fig:figure2}(a) shows that the fast electrons propagate from the cone tip into the dense fuel with relatively little transverse spread, as one would have hoped the mirror scheme to work.  By comparing figures \ref{fig:figure1} and \ref{fig:figure2}, it can also be seen that the fast electrons are very well confined to the semi-ellipsoidal Al guide element.  The magnetic field plot, fig.\ref{fig:figure2}(b), shows that a strong elliptical mirror has been generated at the interface between the Al guide structure and the CH substrate in the cone tip insert.  Clearly it this azimuthal magnetic field structure that is providing the excellent confinement observed in the plot of the fast electron density.  Also note the formation of some magnetic field within the ellipsoid close to the axis, which is due to inhomogeneous propagation of the fast electrons.  These magnetic fields may be an inhibiting factor in achieving the best performance from this configuration, and we return to this point in Sec \ref{sourcerad}.  So the mirror configuration is indeed established, and almost as was expected (i.e. as described in Sec.\ref{ellipmir}).  The energetic coupling performance in this simulation (and all others) was analyzed by examining the energy deposited in a 40$\times$40$\times$40$\mu$m cube centred at (170,100,100)$\mu$m.  Plots of the ion energy density at 20~ps and the ion temperature in the fuel region ($x >$100$\mu$m) are shown in fig. \ref{fig:figure3}  respectively.

Of the 23~kJ of fast electron energy that is injected 4.4~kJ is deposited in the cubic `hot spot' region.  Ion temperatures in the range of 2-4~keV are reached even in the reach close to peak density.  The total energy deposited in the DT fuel is 14.07~kJ.  The rest is either deposited in the cone or lost by fast electrons passing through the far boundaries.  This means there is a nominal coupling efficiency of 19.1\% in this calculation.  This has to be compared to the case without the elliptical mirror in order to assess the benefit that has been derived from the mirror.  The percentage coupling into the same hot-spot is shown for runs CMP1--CMP4 as a function of the varied parameter, $\theta_{div}$, in figure \ref{fig:figure4} below.  

\begin{figure}[H]
\begin{center}
\includegraphics[width = \columnwidth]{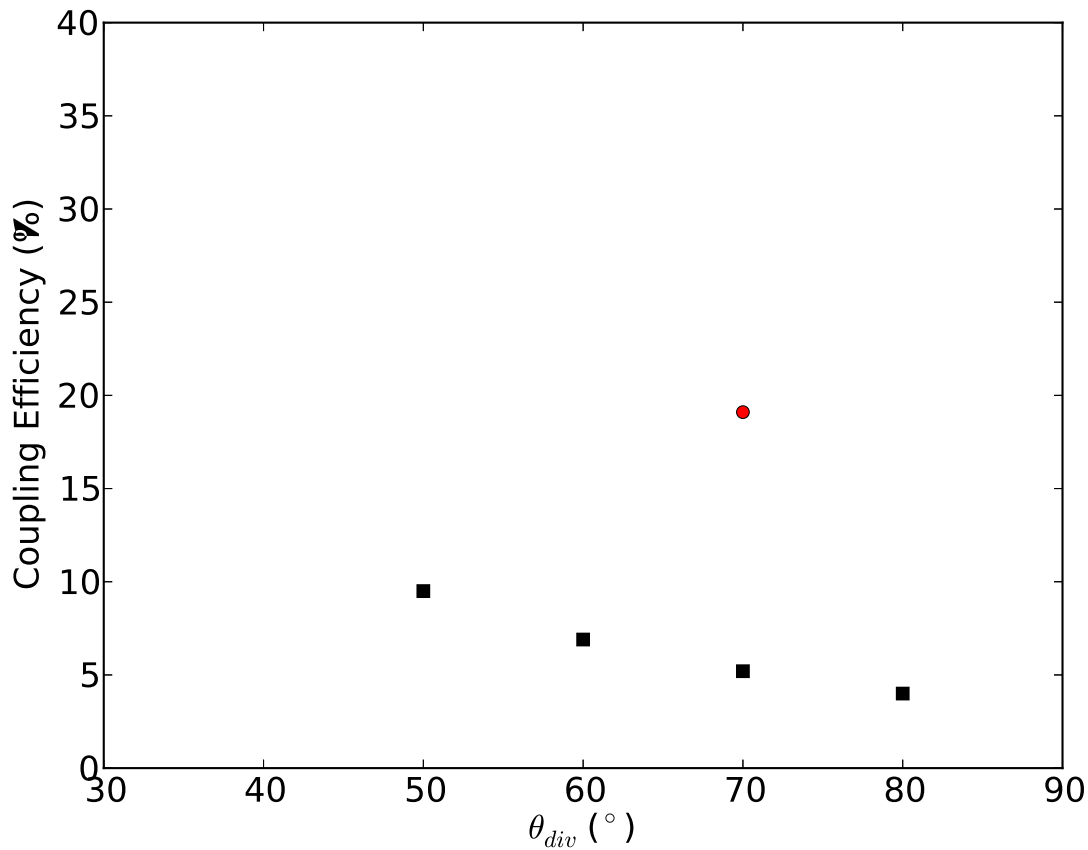} 
\caption{\label{fig:figure4} Coupling efficiency into hot spot in comparator (no mirror) simulations (CMP1--CMP4;black squares), and baseline simulation (A;red circle).}
\end{center}
\end{figure}
The baseline simulation is also shown in figure \ref{fig:figure4}.  Without the elliptical mirror, the coupling efficiency is 5.2\%.  So the elliptical mirror increases the coupling efficiency by a factor of 3.7.  So, the baseline scenario, in which the angular spread of the fast electrons is quite considerable anyway, indicates that the elliptical mirror is highly effective.  Note that the coupling efficiency obtained in the baseline simulation is greater than in all the comparator simulations (CMP1--CMP4).  

\subsection{Effect of Varying Baseline Parameters}
\subsubsection{Fast Electron Divergence Angle}
\begin{figure}[H]
\begin{center}
\includegraphics[width = \columnwidth]{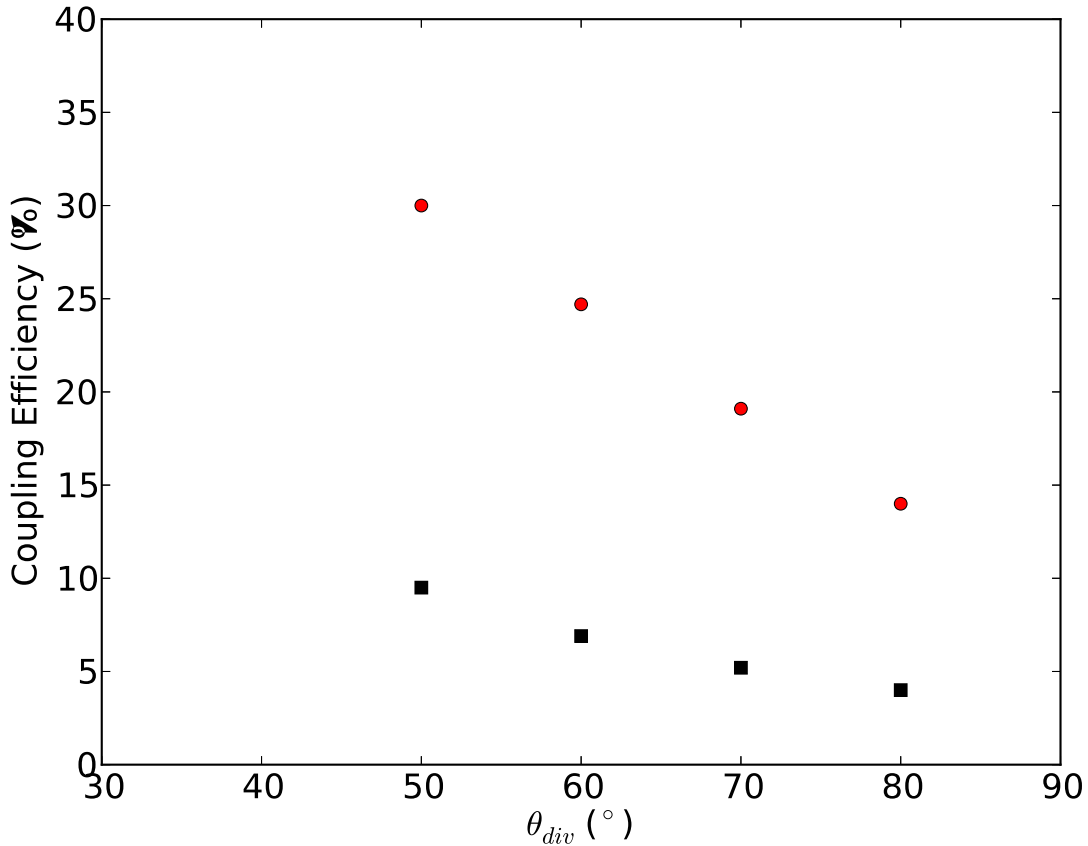} 
\caption{\label{fig:figure5} Coupling efficiency into hot spot in comparator (no mirror) simulations (CMP1--CMP4;black squares), and simulations A--E (red circles).}
\end{center}
\end{figure}

In simulations A--E we have extended the analysis of the baseline simulation, to look at a wider range of fast electron divergence half-angles.  The coupling effiencies that were extracted, including simulation A, are shown in fig. \ref{fig:figure5} alongside the results from CMP1--CMP4.  It can be seen that although the coupling efficiency falls with increasing divergence half-angle, which is entirely expected even with the mirror, the coupling efficiency is always at least 3 times higher with the mirror than without.  We therefore find that the elliptical mirror can be highly beneficial across a range of fast electron divergence angles. 

\subsubsection{Fast Electron Source Radius}
\label{sourcerad}
In simulations F and G we varied the radius of the fast electron source region.  In simulation F, the source radius was 14$\mu$m (compared to 15$\mu$m in A--E).  The total injected fast electron energy was 20.07~kJ and the energy coupled into the hot spot was 4.2~kJ, i.e. a 20.9\% coupling efficiency.  In simulation G, the source radius was 13$\mu$m (compared to 15$\mu$m in A and B).  In G the total fast electron energy was 17.33~kJ, and the energy coupled into the hot spot was 3.07~kJ, i.e. a 17.7\% coupling efficiency.  There is, therefore, some slight dependence on the laser spot radius.  One might expect that some optimization can be performed by tuning the hot-spot radius, because as the hot spot radius is reduced the fast electron source becomes more point-like, and the elliptical mirror configuration is able to focus this into a smaller volume.  These runs show that the improvement that can be gained is only likely to be a little.   If the hot spot radius is reduced too much then there won't be enough energy to achieve ignition, so the extent of this optimization is very limited.

\begin{figure}[H]
\begin{center}
\includegraphics[width = \columnwidth]{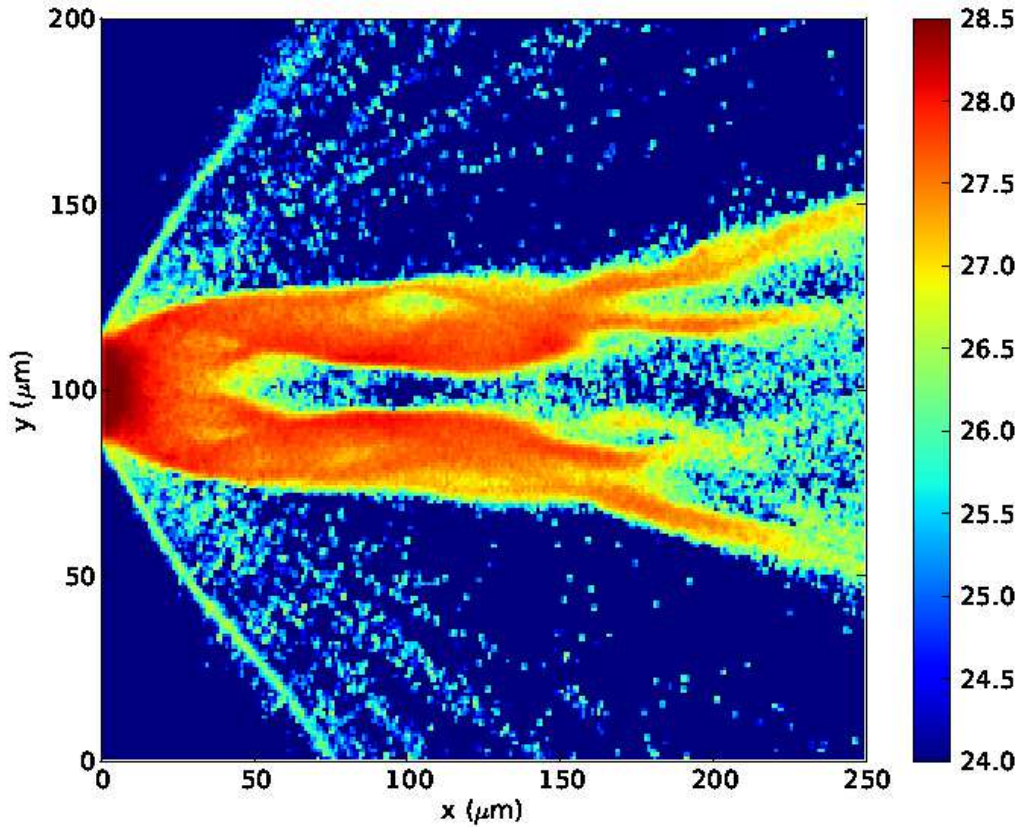} 
\caption{\label{fig:figure5b} (log$_{10}$ fast electron density (in m$^{-3}$) at 4~ps in simulation G in $x$-$y$ mid-plane of simulation box.}
\end{center}
\end{figure}

The poor performance observed in simulation G appears to be due to magnetic fields that develop in the interior of the Al ellipsoid and which produce a strongly annular transport pattern.  This is illustrated by figure \ref{fig:figure5b} below which shows the fast electron density at 4~ps in run G.  The formation of such 'interior' magnetic fields is consistently seen throughout the simulations (see figure \ref{fig:figure2}, where it is clearly visible), and is an obvious consequence of the fast electrons not uniformly filling the ellipsoid as they propagate along its length.  In simulation G these fields have developed much more strongly leading to highly annular transport.  The annular transport leads to a ring-like heating pattern in the DT, hence the coupling efficiency is mainly reduced energy being deposited further out radially.  Clearly this sort of behaviour can occur in this guiding geometry where the ellipsoid is relatively long ($a \gg b$), and future work may have to examine ways to mitigate this problem (although it is currently only observed in a minority of cases).

\subsection{Effect of Alternative Fast Electron Scaling}
In simulations A--G, only a few kJ are coupled into the hot spot.  Partly this is a consequence of the fast electron scaling with laser intensity (pondermotive).  In previous studies of Fast Ignition energy coupling, it has been noted that a ponderomotive energy scaling is not conducive to Fast Ignition.  This is because the fast electron range is approximately given by $\rho{R} =\text{0.6}T\text{(MeV)}$gcm$^{-2}$ \cite{solodov3}, so the ideal mean fast electron energy lies in the range of 1--2~MeV.  The ponderomotive scaling therefore tends to force one to have an unfavourably high ($>$2~MeV) mean fast electron energy in order to inject sufficient energy in under 20~ps.  Some studies of fast electron generation, on the other hand, have observed what might be termed `sub-ponderomotive' fast electron scalings \cite{haines1,kluge1} which, if realizable at full scale, are more conducive to Fast Ignition for the aforementioned reasons.  The matter of fast electron scalings is still a subject of on-going research, and the realization of such a scaling under full Fast Ignition conditions may not be possible.  Nonetheless, we carried out further  simulations to see if the coupling efficiencies are maintained under these different conditions.

The further simulations (H--K)made the following modifications to the baseline simulation: (i) laser intensity was increased to 6$\times$10$^{20}$Wcm$^{-2}$, (ii) $T_f =$0.6$T_p$, (iii) conversion efficiency was decreased to 40\%.  Otherwise the simulation parameters are unchanged.  The particular choice of $T_f =$0.6$T_p$ is taken from the results obtained by Sherlock \cite{sherlock:103101}.  The total injected fast electron energy is 27.65~kJ, so the energetic scale of these simulations is comparable to the others in the series.  The actual value of $T_f$ is 1.95~MeV  The analysis of the energy coupling into the cubic `hot spot' is identical to that done previously.  One extra comparator run (CMP5) was performed.  The simulation results in terms of the coupling efficiency are shown in figure \ref{fig:figure6} below.

\begin{figure}[H]
\begin{center}
\includegraphics[width = \columnwidth]{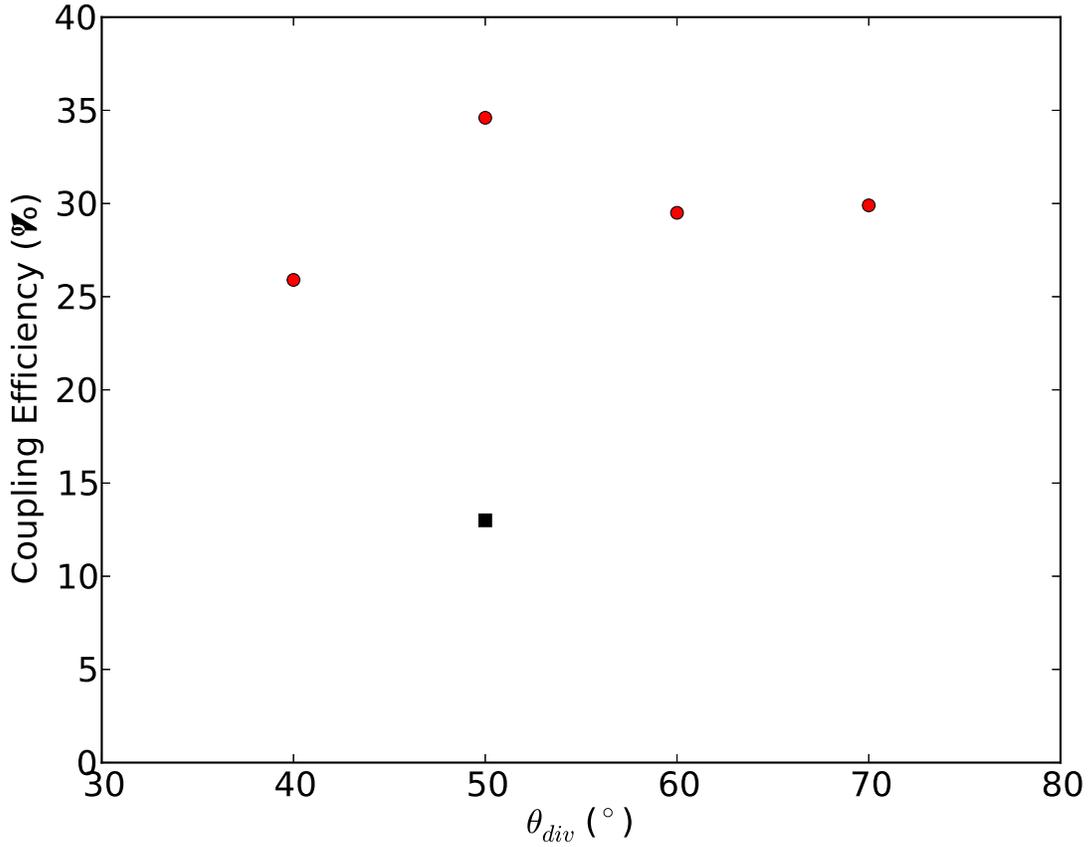} 
\caption{\label{fig:figure6} Coupling efficiency into hot spot in simulations H--K (red circles), and CMP5 (black square).}
\end{center}
\end{figure}

In the equivalent ponderomotive comparator, CMP4, the coupling efficiency was 9.5\%.  As CMP5 has a 13\% coupling efficiency, the effect of lowering $T_f$ has been to improve the coupling efficiency by about 37\%.  The coupling efficiencies with the elliptical mirror are higher than in simulations A--E (see figure \ref{fig:figure5}), but this largely expected from the increase in coupling due to lowering $T_f$.  The variation of coupling efficiency with the fast electron divergence half-angle is flatter overall  From figure \ref{fig:figure6} it can be seen that the elliptical mirror configuration still produces very good coupling efficiencies into a realistically sized hot spot over a wide range of divergence half-angles, and that these coupling efficiencies are better than what is achieved without any guiding scheme by a factor of a few.

\section{Conclusions}
In this paper we have carried out a series of numerical simulations in which we have extended the analysis of the magnetic mirror configuration proposed by Schmitz for FI to conditions which are much closer to full-scale FI.  A number of different aspects of the source-target configuration have been examined, including the fast electron divergence half-angle, spot size, and a sub-ponderomotive fast electron temperature scaling.  We have shown, for multi-kJ conditions, that the elliptical mirror can improve the coupling effiency into the hot spot by a factor of 3--4.  For ponderomotive scaling of the fast electron temperature we have obtained fast electron to hot spot coupling efficiencies of 20--30\% depending on the fast electron divergence half-angle.  In the case of a particular sub-ponderomotive fast electron temperature scaling we observed fast electron to hot spot coupling efficiencies of 25--35\% depending on the fast electron divergence half-angle.  The improvement over the same case without the mirror configuration was about 2.5--3.  Although considerable improvements in coupling efficiency are easily demonstrated, as shown throughout this study, there is the potential problem of the annular transport mode (discussed in Sec. \ref{sourcerad}).  This may be problematic, but, given that this was only problematic in {it one} case in this study, we believe this to be a minor concern.

We therefore conclude that Schmitz's elliptical mirror scheme has considerable potential for improving the prospects of Fast Ignition, as a factor of 3--4 improvement in coupling efficiency is quite substantial.  The elliptical mirror design used in this study is not necessarily optimal, and we have assumed that it is one of the better choices of geometry for the reasons given in Sec.\ref{ellipmir}.  Future work is needed to examine the optimization of the elliptical mirror scheme, or whether some modified or alternative geometry can further improve coupling efficiency into the hot spot.

\begin{acknowledgements}
This work was supported by the European Research Council's  STRUCMAGFAST grant (ERC-StG-2012).  APLR and HS are grateful for the use of computing resources provided by STFC's Scientific Computing Department.
\end{acknowledgements}


%

\begin{thebibliography}{31}%
\makeatletter
\providecommand \@ifxundefined [1]{%
 \@ifx{#1\undefined}
}%
\providecommand \@ifnum [1]{%
 \ifnum #1\expandafter \@firstoftwo
 \else \expandafter \@secondoftwo
 \fi
}%
\providecommand \@ifx [1]{%
 \ifx #1\expandafter \@firstoftwo
 \else \expandafter \@secondoftwo
 \fi
}%
\providecommand \natexlab [1]{#1}%
\providecommand \enquote  [1]{``#1''}%
\providecommand \bibnamefont  [1]{#1}%
\providecommand \bibfnamefont [1]{#1}%
\providecommand \citenamefont [1]{#1}%
\providecommand \href@noop [0]{\@secondoftwo}%
\providecommand \href [0]{\begingroup \@sanitize@url \@href}%
\providecommand \@href[1]{\@@startlink{#1}\@@href}%
\providecommand \@@href[1]{\endgroup#1\@@endlink}%
\providecommand \@sanitize@url [0]{\catcode `\\12\catcode `\$12\catcode
  `\&12\catcode `\#12\catcode `\^12\catcode `\_12\catcode `\%12\relax}%
\providecommand \@@startlink[1]{}%
\providecommand \@@endlink[0]{}%
\providecommand \url  [0]{\begingroup\@sanitize@url \@url }%
\providecommand \@url [1]{\endgroup\@href {#1}{\urlprefix }}%
\providecommand \urlprefix  [0]{URL }%
\providecommand \Eprint [0]{\href }%
\providecommand \doibase [0]{http://dx.doi.org/}%
\providecommand \selectlanguage [0]{\@gobble}%
\providecommand \bibinfo  [0]{\@secondoftwo}%
\providecommand \bibfield  [0]{\@secondoftwo}%
\providecommand \translation [1]{[#1]}%
\providecommand \BibitemOpen [0]{}%
\providecommand \bibitemStop [0]{}%
\providecommand \bibitemNoStop [0]{.\EOS\space}%
\providecommand \EOS [0]{\spacefactor3000\relax}%
\providecommand \BibitemShut  [1]{\csname bibitem#1\endcsname}%
\let\auto@bib@innerbib\@empty
\bibitem [{\citenamefont {H.Schmitz}\ \emph {et~al.}(2012)\citenamefont
  {H.Schmitz} \emph {et~al.}}]{schmitz1}%
  \BibitemOpen
  \bibfield  {author} {\bibinfo {author} {\bibnamefont {H.Schmitz}} \emph
  {et~al.},\ }\href@noop {} {\bibfield  {journal} {\bibinfo  {journal} {Plasma
  Phys.Control.Fusion}\ }\textbf {\bibinfo {volume} {54}},\ \bibinfo {pages}
  {085016} (\bibinfo {year} {2012})}\BibitemShut {NoStop}%
\bibitem [{\citenamefont {A.P.L.Robinson}\ and\ \citenamefont
  {M.Sherlock}(2007)}]{robinson1}%
  \BibitemOpen
  \bibfield  {author} {\bibinfo {author} {\bibnamefont {A.P.L.Robinson}}\ and\
  \bibinfo {author} {\bibnamefont {M.Sherlock}},\ }\href@noop {} {\bibfield
  {journal} {\bibinfo  {journal} {Phys.Plasmas}\ }\textbf {\bibinfo {volume}
  {14}},\ \bibinfo {pages} {083105} (\bibinfo {year} {2007})}\BibitemShut
  {NoStop}%
\bibitem [{\citenamefont {M.Tabak}\ \emph {et~al.}(1994)\citenamefont {M.Tabak}
  \emph {et~al.}}]{tabak1}%
  \BibitemOpen
  \bibfield  {author} {\bibinfo {author} {\bibnamefont {M.Tabak}} \emph
  {et~al.},\ }\href@noop {} {\bibfield  {journal} {\bibinfo  {journal}
  {Phys.Plasmas}\ }\textbf {\bibinfo {volume} {1}} (\bibinfo {year}
  {1994})}\BibitemShut {NoStop}%
\bibitem [{\citenamefont {M.Tabak}\ \emph {et~al.}(2005)\citenamefont {M.Tabak}
  \emph {et~al.}}]{tabak2}%
  \BibitemOpen
  \bibfield  {author} {\bibinfo {author} {\bibnamefont {M.Tabak}} \emph
  {et~al.},\ }\href@noop {} {\bibfield  {journal} {\bibinfo  {journal}
  {Phys.Plasmas}\ }\textbf {\bibinfo {volume} {{\bf 12}}},\ \bibinfo {pages}
  {057305} (\bibinfo {year} {2005})}\BibitemShut {NoStop}%
\bibitem [{\citenamefont {S.Atzeni}\ \emph {et~al.}(2007)\citenamefont
  {S.Atzeni}, \citenamefont {A.Schiavi},\ and\ \citenamefont
  {C.Bellei}}]{atzeni1}%
  \BibitemOpen
  \bibfield  {author} {\bibinfo {author} {\bibnamefont {S.Atzeni}}, \bibinfo
  {author} {\bibnamefont {A.Schiavi}}, \ and\ \bibinfo {author} {\bibnamefont
  {C.Bellei}},\ }\href@noop {} {\bibfield  {journal} {\bibinfo  {journal}
  {Phys.Plasmas}\ }\textbf {\bibinfo {volume} {14}},\ \bibinfo {pages} {052702}
  (\bibinfo {year} {2007})}\BibitemShut {NoStop}%
\bibitem [{\citenamefont {J.J.Honrubia}\ and\ \citenamefont {ter
  Vehn}(2009)}]{honrubia1}%
  \BibitemOpen
  \bibfield  {author} {\bibinfo {author} {\bibnamefont {J.J.Honrubia}}\ and\
  \bibinfo {author} {\bibfnamefont {J.}~\bibnamefont {ter Vehn}},\ }\href@noop
  {} {\bibfield  {journal} {\bibinfo  {journal} {Plasma Phys.Control.Fusion}\
  }\textbf {\bibinfo {volume} {51}},\ \bibinfo {pages} {014008} (\bibinfo
  {year} {2009})}\BibitemShut {NoStop}%
\bibitem [{\citenamefont {T.Johzaki}\ \emph {et~al.}(2011)\citenamefont
  {T.Johzaki} \emph {et~al.}}]{johzaki1}%
  \BibitemOpen
  \bibfield  {author} {\bibinfo {author} {\bibnamefont {T.Johzaki}} \emph
  {et~al.},\ }\href@noop {} {\bibfield  {journal} {\bibinfo  {journal}
  {Nucl.Fusion}\ }\textbf {\bibinfo {volume} {51}},\ \bibinfo {pages} {073022}
  (\bibinfo {year} {2011})}\BibitemShut {NoStop}%
\bibitem [{\citenamefont {T.Johzaki}\ \emph {et~al.}(2009)\citenamefont
  {T.Johzaki} \emph {et~al.}}]{johzaki2}%
  \BibitemOpen
  \bibfield  {author} {\bibinfo {author} {\bibnamefont {T.Johzaki}} \emph
  {et~al.},\ }\href@noop {} {\bibfield  {journal} {\bibinfo  {journal}
  {Phys.Plasmas}\ }\textbf {\bibinfo {volume} {16}},\ \bibinfo {pages} {062706}
  (\bibinfo {year} {2009})}\BibitemShut {NoStop}%
\bibitem [{\citenamefont {A.A.Solodov}\ \emph {et~al.}(2008)\citenamefont
  {A.A.Solodov} \emph {et~al.}}]{solodov1}%
  \BibitemOpen
  \bibfield  {author} {\bibinfo {author} {\bibnamefont {A.A.Solodov}} \emph
  {et~al.},\ }\href@noop {} {\bibfield  {journal} {\bibinfo  {journal}
  {Phys.Plasmas}\ }\textbf {\bibinfo {volume} {15}},\ \bibinfo {pages} {112702}
  (\bibinfo {year} {2008})}\BibitemShut {NoStop}%
\bibitem [{\citenamefont {A.A.Solodov}\ \emph {et~al.}(2007)\citenamefont
  {A.A.Solodov} \emph {et~al.}}]{solodov2}%
  \BibitemOpen
  \bibfield  {author} {\bibinfo {author} {\bibnamefont {A.A.Solodov}} \emph
  {et~al.},\ }\href@noop {} {\bibfield  {journal} {\bibinfo  {journal}
  {Phys.Plasmas}\ }\textbf {\bibinfo {volume} {14}},\ \bibinfo {pages} {062701}
  (\bibinfo {year} {2007})}\BibitemShut {NoStop}%
\bibitem [{\citenamefont {Atzeni}\ \emph {et~al.}(2008)\citenamefont {Atzeni},
  \citenamefont {Schiavi}, \citenamefont {Honrubia}, \citenamefont {Ribeyre},
  \citenamefont {Schurtz}, \citenamefont {Nicolai}, \citenamefont
  {Olazabal-Loume}, \citenamefont {Bellei}, \citenamefont {Evans},\ and\
  \citenamefont {Davies}}]{atzeni3}%
  \BibitemOpen
  \bibfield  {author} {\bibinfo {author} {\bibfnamefont {S.}~\bibnamefont
  {Atzeni}}, \bibinfo {author} {\bibfnamefont {A.}~\bibnamefont {Schiavi}},
  \bibinfo {author} {\bibfnamefont {J.~J.}\ \bibnamefont {Honrubia}}, \bibinfo
  {author} {\bibfnamefont {X.}~\bibnamefont {Ribeyre}}, \bibinfo {author}
  {\bibfnamefont {G.}~\bibnamefont {Schurtz}}, \bibinfo {author} {\bibfnamefont
  {P.}~\bibnamefont {Nicolai}}, \bibinfo {author} {\bibfnamefont
  {M.}~\bibnamefont {Olazabal-Loume}}, \bibinfo {author} {\bibfnamefont
  {C.}~\bibnamefont {Bellei}}, \bibinfo {author} {\bibfnamefont {R.~G.}\
  \bibnamefont {Evans}}, \ and\ \bibinfo {author} {\bibfnamefont {J.~R.}\
  \bibnamefont {Davies}},\ }\href {\doibase 10.1063/1.2895447} {\bibfield
  {journal} {\bibinfo  {journal} {Physics of Plasmas}\ }\textbf {\bibinfo
  {volume} {15}},\ \bibinfo {eid} {056311} (\bibinfo {year}
  {2008})}\BibitemShut {NoStop}%
\bibitem [{\citenamefont {P.Nicolai}\ \emph {et~al.}(2011)\citenamefont
  {P.Nicolai} \emph {et~al.}}]{nicolai1}%
  \BibitemOpen
  \bibfield  {author} {\bibinfo {author} {\bibnamefont {P.Nicolai}} \emph
  {et~al.},\ }\href@noop {} {\bibfield  {journal} {\bibinfo  {journal}
  {Phys.Rev. E}\ }\textbf {\bibinfo {volume} {84}},\ \bibinfo {pages} {016402}
  (\bibinfo {year} {2011})}\BibitemShut {NoStop}%
\bibitem [{\citenamefont {K.L.Lancaster}\ \emph {et~al.}(2007)\citenamefont
  {K.L.Lancaster} \emph {et~al.}}]{lancaster1}%
  \BibitemOpen
  \bibfield  {author} {\bibinfo {author} {\bibnamefont {K.L.Lancaster}} \emph
  {et~al.},\ }\href@noop {} {\bibfield  {journal} {\bibinfo  {journal}
  {Phys.Rev.Lett.}\ }\textbf {\bibinfo {volume} {98}},\ \bibinfo {pages}
  {125002} (\bibinfo {year} {2007})}\BibitemShut {NoStop}%
\bibitem [{\citenamefont {J.S.Green}\ \emph {et~al.}(2008)\citenamefont
  {J.S.Green} \emph {et~al.}}]{green1}%
  \BibitemOpen
  \bibfield  {author} {\bibinfo {author} {\bibnamefont {J.S.Green}} \emph
  {et~al.},\ }\href@noop {} {\bibfield  {journal} {\bibinfo  {journal}
  {Phys.Rev.Lett.}\ }\textbf {\bibinfo {volume} {100}},\ \bibinfo {pages}
  {015003} (\bibinfo {year} {2008})}\BibitemShut {NoStop}%
\bibitem [{\citenamefont {A.Debayle}(2010)}]{debayle1}%
  \BibitemOpen
  \bibfield  {author} {\bibinfo {author} {\bibnamefont {A.Debayle}},\
  }\href@noop {} {\bibfield  {journal} {\bibinfo  {journal} {Phys.Rev.E}\
  }\textbf {\bibinfo {volume} {{\bf 82}}},\ \bibinfo {pages} {036405} (\bibinfo
  {year} {2010})}\BibitemShut {NoStop}%
\bibitem [{\citenamefont {R.Stephens}\ \emph {et~al.}(2004)\citenamefont
  {R.Stephens} \emph {et~al.}}]{stephens1}%
  \BibitemOpen
  \bibfield  {author} {\bibinfo {author} {\bibnamefont {R.Stephens}} \emph
  {et~al.},\ }\href@noop {} {\bibfield  {journal} {\bibinfo  {journal}
  {Phys.Rev.E}\ }\textbf {\bibinfo {volume} {69}},\ \bibinfo {pages} {066414}
  (\bibinfo {year} {2004})}\BibitemShut {NoStop}%
\bibitem [{\citenamefont {P.A.Norreys}\ \emph {et~al.}(2009)\citenamefont
  {P.A.Norreys} \emph {et~al.}}]{norreys1}%
  \BibitemOpen
  \bibfield  {author} {\bibinfo {author} {\bibnamefont {P.A.Norreys}} \emph
  {et~al.},\ }\href@noop {} {\bibfield  {journal} {\bibinfo  {journal} {Nucl.
  Fusion}\ }\textbf {\bibinfo {volume} {49}},\ \bibinfo {pages} {104023}
  (\bibinfo {year} {2009})}\BibitemShut {NoStop}%
\bibitem [{\citenamefont {A.R.Bell}\ and\ \citenamefont
  {R.J.Kingham}(2003)}]{bell1}%
  \BibitemOpen
  \bibfield  {author} {\bibinfo {author} {\bibnamefont {A.R.Bell}}\ and\
  \bibinfo {author} {\bibnamefont {R.J.Kingham}},\ }\href@noop {} {\bibfield
  {journal} {\bibinfo  {journal} {Phys.Rev.Lett.}\ }\textbf {\bibinfo {volume}
  {91}},\ \bibinfo {pages} {035003} (\bibinfo {year} {2003})}\BibitemShut
  {NoStop}%
\bibitem [{\citenamefont {R.B.Campbell}\ \emph {et~al.}(2003)\citenamefont
  {R.B.Campbell} \emph {et~al.}}]{campbell1}%
  \BibitemOpen
  \bibfield  {author} {\bibinfo {author} {\bibnamefont {R.B.Campbell}} \emph
  {et~al.},\ }\href@noop {} {\bibfield  {journal} {\bibinfo  {journal}
  {Phys.Plasmas}\ }\textbf {\bibinfo {volume} {10}},\ \bibinfo {pages} {4169}
  (\bibinfo {year} {2003})}\BibitemShut {NoStop}%
\bibitem [{\citenamefont {T.Nakamura}\ \emph {et~al.}(2007)\citenamefont
  {T.Nakamura} \emph {et~al.}}]{nakamura1}%
  \BibitemOpen
  \bibfield  {author} {\bibinfo {author} {\bibnamefont {T.Nakamura}} \emph
  {et~al.},\ }\href@noop {} {\bibfield  {journal} {\bibinfo  {journal}
  {Phys.Plasmas}\ }\textbf {\bibinfo {volume} {14}},\ \bibinfo {pages} {103105}
  (\bibinfo {year} {2007})}\BibitemShut {NoStop}%
\bibitem [{\citenamefont {Kodama}\ \emph {et~al.}(2004)\citenamefont {Kodama}
  \emph {et~al.}}]{kodama1}%
  \BibitemOpen
  \bibfield  {author} {\bibinfo {author} {\bibfnamefont {R.}~\bibnamefont
  {Kodama}} \emph {et~al.},\ }\href@noop {} {\bibfield  {journal} {\bibinfo
  {journal} {Nature}\ }\textbf {\bibinfo {volume} {432}},\ \bibinfo {pages}
  {1005} (\bibinfo {year} {2004})}\BibitemShut {NoStop}%
\bibitem [{\citenamefont {A.P.L.Robinson}\ \emph {et~al.}(2008)\citenamefont
  {A.P.L.Robinson}, \citenamefont {M.Sherlock},\ and\ \citenamefont
  {P.A.Norreys}}]{robinson3}%
  \BibitemOpen
  \bibfield  {author} {\bibinfo {author} {\bibnamefont {A.P.L.Robinson}},
  \bibinfo {author} {\bibnamefont {M.Sherlock}}, \ and\ \bibinfo {author}
  {\bibnamefont {P.A.Norreys}},\ }\href@noop {} {\bibfield  {journal} {\bibinfo
   {journal} {Phys.Rev.Lett.}\ }\textbf {\bibinfo {volume} {100}},\ \bibinfo
  {pages} {025002} (\bibinfo {year} {2008})}\BibitemShut {NoStop}%
\bibitem [{\citenamefont {R.H.H.Scott}\ \emph {et~al.}(2012)\citenamefont
  {R.H.H.Scott} \emph {et~al.}}]{scott1}%
  \BibitemOpen
  \bibfield  {author} {\bibinfo {author} {\bibnamefont {R.H.H.Scott}} \emph
  {et~al.},\ }\href@noop {} {\bibfield  {journal} {\bibinfo  {journal}
  {Phys.Rev.Lett.}\ }\textbf {\bibinfo {volume} {109}},\ \bibinfo {pages}
  {015001} (\bibinfo {year} {2012})}\BibitemShut {NoStop}%
\bibitem [{\citenamefont {S.Kar}\ \emph {et~al.}(2009)\citenamefont {S.Kar}
  \emph {et~al.}}]{kar1}%
  \BibitemOpen
  \bibfield  {author} {\bibinfo {author} {\bibnamefont {S.Kar}} \emph
  {et~al.},\ }\href@noop {} {\bibfield  {journal} {\bibinfo  {journal}
  {Phys.Rev.Lett.}\ }\textbf {\bibinfo {volume} {102}},\ \bibinfo {pages}
  {055001} (\bibinfo {year} {2009})}\BibitemShut {NoStop}%
\bibitem [{\citenamefont {B.Ramakrishna}\ \emph {et~al.}(2010)\citenamefont
  {B.Ramakrishna} \emph {et~al.}}]{ramak1}%
  \BibitemOpen
  \bibfield  {author} {\bibinfo {author} {\bibnamefont {B.Ramakrishna}} \emph
  {et~al.},\ }\href@noop {} {\bibfield  {journal} {\bibinfo  {journal}
  {Phys.Rev.Lett.}\ }\textbf {\bibinfo {volume} {105}},\ \bibinfo {pages}
  {135001} (\bibinfo {year} {2010})}\BibitemShut {NoStop}%
\bibitem [{\citenamefont {J.R.Davies}(2002)}]{jrd1}%
  \BibitemOpen
  \bibfield  {author} {\bibinfo {author} {\bibnamefont {J.R.Davies}},\
  }\href@noop {} {\bibfield  {journal} {\bibinfo  {journal} {Phys.Rev.E}\
  }\textbf {\bibinfo {volume} {{\bf 65}}},\ \bibinfo {pages} {026407} (\bibinfo
  {year} {2002})}\BibitemShut {NoStop}%
\bibitem [{\citenamefont {A.P.L.Robinson}\ \emph {et~al.}(2012)\citenamefont
  {A.P.L.Robinson}, \citenamefont {M.H.Key},\ and\ \citenamefont
  {M.Tabak}}]{robinson2}%
  \BibitemOpen
  \bibfield  {author} {\bibinfo {author} {\bibnamefont {A.P.L.Robinson}},
  \bibinfo {author} {\bibnamefont {M.H.Key}}, \ and\ \bibinfo {author}
  {\bibnamefont {M.Tabak}},\ }\href@noop {} {\bibfield  {journal} {\bibinfo
  {journal} {Phys.Rev.Lett.}\ }\textbf {\bibinfo {volume} {108}},\ \bibinfo
  {pages} {125004} (\bibinfo {year} {2012})}\BibitemShut {NoStop}%
\bibitem [{\citenamefont {A.A.Solodov}\ and\ \citenamefont
  {R.Betti}(2008)}]{solodov3}%
  \BibitemOpen
  \bibfield  {author} {\bibinfo {author} {\bibnamefont {A.A.Solodov}}\ and\
  \bibinfo {author} {\bibnamefont {R.Betti}},\ }\href@noop {} {\bibfield
  {journal} {\bibinfo  {journal} {Phys.Plasmas}\ }\textbf {\bibinfo {volume}
  {15}},\ \bibinfo {pages} {042707} (\bibinfo {year} {2008})}\BibitemShut
  {NoStop}%
\bibitem [{\citenamefont {M.G.Haines}\ \emph {et~al.}(2009)\citenamefont
  {M.G.Haines} \emph {et~al.}}]{haines1}%
  \BibitemOpen
  \bibfield  {author} {\bibinfo {author} {\bibnamefont {M.G.Haines}} \emph
  {et~al.},\ }\href@noop {} {\bibfield  {journal} {\bibinfo  {journal}
  {Phys.Rev.Lett.}\ }\textbf {\bibinfo {volume} {102}},\ \bibinfo {pages}
  {0450089} (\bibinfo {year} {2009})}\BibitemShut {NoStop}%
\bibitem [{\citenamefont {Kluge}\ \emph {et~al.}(2011)\citenamefont {Kluge},
  \citenamefont {Cowan}, \citenamefont {Debus}, \citenamefont {Schramm},
  \citenamefont {Zeil},\ and\ \citenamefont {Bussmann}}]{kluge1}%
  \BibitemOpen
  \bibfield  {author} {\bibinfo {author} {\bibfnamefont {T.}~\bibnamefont
  {Kluge}}, \bibinfo {author} {\bibfnamefont {T.}~\bibnamefont {Cowan}},
  \bibinfo {author} {\bibfnamefont {A.}~\bibnamefont {Debus}}, \bibinfo
  {author} {\bibfnamefont {U.}~\bibnamefont {Schramm}}, \bibinfo {author}
  {\bibfnamefont {K.}~\bibnamefont {Zeil}}, \ and\ \bibinfo {author}
  {\bibfnamefont {M.}~\bibnamefont {Bussmann}},\ }\href {\doibase
  10.1103/PhysRevLett.107.205003} {\bibfield  {journal} {\bibinfo  {journal}
  {Phys. Rev. Lett.}\ }\textbf {\bibinfo {volume} {107}},\ \bibinfo {pages}
  {205003} (\bibinfo {year} {2011})}\BibitemShut {NoStop}%
\bibitem [{\citenamefont {Sherlock}(2009)}]{sherlock:103101}%
  \BibitemOpen
  \bibfield  {author} {\bibinfo {author} {\bibfnamefont {M.}~\bibnamefont
  {Sherlock}},\ }\href {\doibase 10.1063/1.3240341} {\bibfield  {journal}
  {\bibinfo  {journal} {Physics of Plasmas}\ }\textbf {\bibinfo {volume}
  {16}},\ \bibinfo {eid} {103101} (\bibinfo {year} {2009})}\BibitemShut
  {NoStop}%
\end{thebibliography}
\end{document}